\documentclass{sig-alternate}
\usepackage{times}
\usepackage{helvet}
\usepackage{courier}
\usepackage{tabu}
\usepackage{booktabs}
\usepackage{color}
\usepackage[usenames,dvipsnames]{xcolor}
\usepackage{enumitem}
\usepackage{graphicx}
\usepackage{xspace}
\usepackage{mdwlist}
\usepackage{amsmath}
\usepackage{lscape}
\usepackage[small]{caption}
\usepackage{subcaption}
\usepackage{multirow}

\newcommand{\danai}[1]{{\small\color{Rhodamine}{\bf dk: #1}}}
\newcommand{\pnb}[1]{{\small\color{blue}{\bf pnb: #1}}}

\newcommand{\reviewer}[1]{{\small\color{red}{\bf reviewer: #1}}}

\renewcommand{\reviewer}[1]{}
\renewcommand{\danai}[1]{}
\renewcommand{\pnb}[1]{}

\begin{document}

\newcommand{\SH}{Sandy Hook\xspace}
\newcommand{\SHshort}{S.H.\xspace}

\newcommand{\news}{\textsc{News}\xspace} 
\newcommand{\nonNews}{\textsc{GC-Debate}\xspace} 
\newcommand{\newsT}{{\large \textsc{News}}\xspace} 
\newcommand{\nonNewsT}{{\large \textsc{GC-Debate}}\xspace}
\newcommand{\nonNewsCommon}{\textsc{Non-News-CommonUsers}\xspace}

\newcommand{\GMO}{\textsc{GMfood}\xspace} 

\newcommand{\reminder}[1]{{\color{red} (** #1 **)}}
\newcommand{\hide}[1]{}

\newtheorem{observation}{\textbf{Observation}}

\newcommand{\Pmat}{\mathbf{P}} 
\newcommand{\Qmat}{\mathbf{Q}}
\newcommand{\Imat}{\mathbf{I}}

\newcommand{\control}[1]{{\color{blue}{#1}}}
\newcommand{\rights}[1]{{\color{red}{#1}}}
\newcommand{\balanced}[1]{{\color{gray}{#1}}}

\title{Events and Controversies: \\Influences of a Shocking News Event on Information Seeking}

%
\numberofauthors{3}
\author{
\alignauthor
Danai Koutra\titlenote{Work done during an internship at Microsoft Research.}\\
       \affaddr{Carnegie Mellon University}\\
       \affaddr{5000 Forbes Avenue}\\
       \affaddr{Pittsburgh, PA}\\
       \email{danai@cs.cmu.com}
\alignauthor
Paul Bennett \\
       \affaddr{Microsoft Research}\\
       \affaddr{One Microsoft Way}\\
       \affaddr{Redmond, WA}\\
       \email{pauben@microsoft.com}
\alignauthor
 Eric Horvitz\\
       \affaddr{Microsoft Research}\\
       \affaddr{One Microsoft Way}\\
       \affaddr{Redmond, WA}\\
       \email{horvitz@microsoft.com}
}
\maketitle


\begin{abstract}
\begin{quote}
It has been suggested that online search and retrieval contributes
to the intellectual isolation of users within their preexisting ideologies,
where people's prior views are strengthened and alternative
viewpoints are infrequently encountered. This so-called ``filter
bubble'' phenomenon has been called out as especially
detrimental when it comes to dialog among people on controversial,
emotionally charged topics, such as the labeling of genetically
modified food, the right to bear arms, the death penalty, and online
privacy. 
We seek to identify and study information-seeking behavior 
and access to alternative versus reinforcing viewpoints following shocking, emotional,
and large-scale news events.  We choose for a case study to analyze search and browsing on
gun control/rights, a strongly polarizing topic for both citizens
and leaders of the United States. We study the period of
time preceding and following a mass shooting to understand how
its occurrence, follow-on discussions, and debate may have been
linked to changes in the patterns of searching and browsing. We employ information-theoretic measures to quantify the diversity of
Web domains of interest to users and understand the browsing patterns of users. We use these measures to characterize the influence of news events on these web search and browsing patterns.
\end{quote}
\end{abstract}

\section{Introduction}

How do people navigate webpages on polarizing topics? 
Are they isolated in their echo chambers? 
Do shocking news events burst their ideological bubbles and make them more likely to seek information on opposing viewpoints? 
These are the key questions we investigate in this work.

With advances in personalization methods, search engines and recommendation systems increasingly adjust results to users' preferences, as inferred from their past searches and choices. In addition, users often input biased queries \cite{YomTovDG13}, which reflect their own positions, while personalized results have the potential to just reinforce these opinions, acting as echo chambers. 
As a result, according to several recent studies \cite{YomTovDG13,GarrettR11,ResnickGKMS13}, the users remain within informational bubbles. The phenomenon is sometimes referred to as the ``filter bubble'' effect, where users get exposed only to opinions that align with their current views. This effect, where the ``web world'' does not reflect the richness of views in the ``real world'' may be exacerbated for polarizing topics.  We take as polarizing or controversial topics those linked to opposing perspectives, such as 
abortion, gun control vs.\ rights, labeling of genetically modified food, and death penalty. 

\hide{people using the web are limited in their own informational bubbles, also referred to as ``filter bubbles'', where they get exposed only to agreeable opinions. This effect, where the ``web world'' does not reflect the variety of views existing in the ``real world'', is exacerbated when polarizing topics are concerned. In short, polarizing or controversial topics are the issues for which multiple contradictory views exist, and usually the truth lies somewhere in the middle. The list of the most controversial topics comprises abortion, gun control and rights, labeling of genetically modified food, death penalty, privacy in social media, college entrance exams, climate change, human cloning.}

\begin{figure*}[t!]
  \centering
        \includegraphics[width=\textwidth]{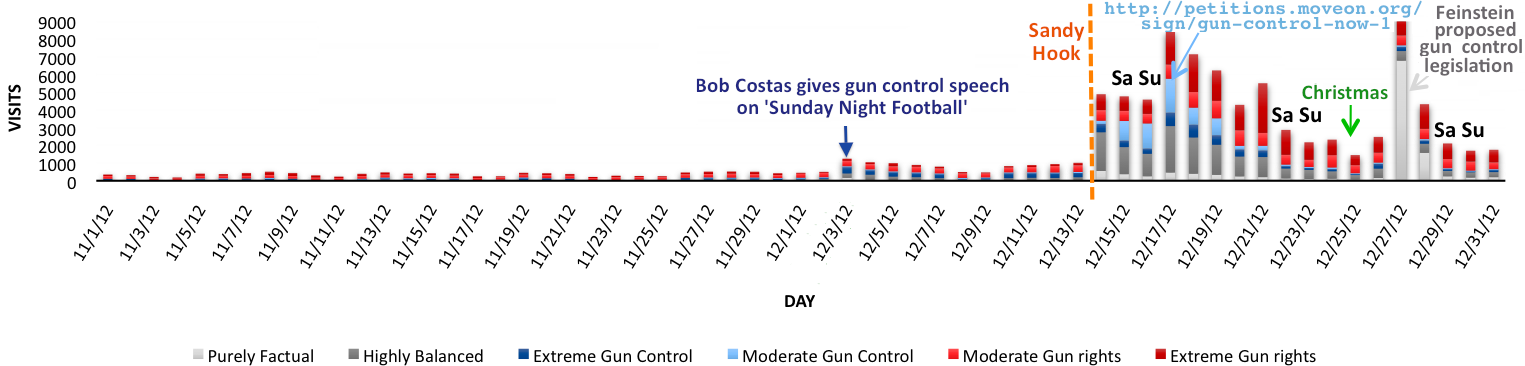}
    
    \caption{ 
    Number of visits to gun control/rights related webpages over time (November-December 2012). The colors correspond to webpages categories: gray for factual and balanced pages; blue for pages supporting gun control; and red for pages supporting gun rights. The categories and the labeling process are described in the Appendix and Sec.~\ref{sec:annotation} respectively.}
    \label{fig:crownJewel}
\end{figure*}

\enlargethispage{- \baselineskip}

In order to understand the users' information seeking behaviors on polarizing issues, we focus on a highly controversial topic 
in the US: gun control and rights. At one end of the spectrum, extreme gun rights supporters argue an interpretation of the 2nd Amendment to the US Constitution that would prohibit any regulation of firearms. On the other side 
of the spectrum,
extreme gun control supporters advocate the total ban of any private citizen ownership of firearms. In addition to these two extreme opinions, there exist multiple variations which lie between them ({\em e.g.}, more background checks, ban of fully automatic firearms). For our study we use web browser toolbar logs from November and December 2012, and primarily consider two time periods: before and after the Sandy Hook Elementary School shootings (\SHshort) in Newtown, Connecticut (December 14th), an event with broad news coverage and nationwide impact. 
The event clearly had considerable influence on information seeking about gun control related topics as signified by the increased user activity in the days following the event (see Fig.~\ref{fig:crownJewel}). Both the first big spike in the figure, which corresponds to visits to on-topic websites 
on the day of the shootings, and other important spikes have been annotated. The effect on the quantity of information seeking is indisputable; so our focus is {\em not} on the increase in user activity, {\em but} on whether (and how) the event changed the {\em type} of activity. 

For the following analysis, we use raw web browser toolbar visitation logs from a popular commercial web browser where users have given consent to logging all non-https URLs from URLs visited from search and those reached by direct entry or browsing. 
By employing techniques such as a two-step random walk on the query-click graph \cite{craswell-szummer:sigir-2007} and 
white-list and keyword-based classifiers,
we extract from this broad set of visitations, a large-scale dataset of user interaction data that is relevant to the gun debate, constituting about 61K users visiting 297K on-topic websites.  Treating this as a sample of a broad spectrum of information seeking behaviors across the US, we then prioritize labeling of the data to label all on-topic sites that at least two users have visited - including over 1,500 websites.

Using this labeled data, we first present evidence that websites are polarized with respect to individual topics in terms of their webpage content (Sec.~\ref{sec:domainDiversity}), and based on this observation, as a simple form of classification, we propagate the manual labels within those web domains that lack diversity.  This enables us to use a large number of labeled sites as the basis for our further studies. 
 Then, 
 we evaluate the diversity of the users and investigate to what extent ideological bubbles exist before and after the shootings (Sec.~\ref{sec:userDiversity}). 
 Finally, we explore the click trails of the users to understand 
 how people transition among webpages of opposing views, and how news about the shootings influences such transitions (Sec.~\ref{sec:change_in_transitions}). 
\pnb{Somewhere in the intro we should sketch some kinds of conclusions.}

 


\hide{
We organize the paper as follows: We start with the data extraction and labeling methodology (Sec.~\ref{sec:gun_data}, \ref{sec:domainDiversity}). 
Next we focus on the users' diversity, the evolution of communities (Sec.~\ref{sec:userDiversity}), the transitions among webpages and polarity shift after the shooting  (Sec.~\ref{sec:change_in_transitions}). We end with related work and conclusions (Sec.~\ref{sec:relatedWork}, \ref{sec:conclusions}).
}

\section{Related Work}
\label{sec:relatedWork}
We first place our work in the context of related research, which includes
 studies on political controversies, conjectures about the so-called filter bubble, and the temporal evolution of knowledge.

\textbf{Political Controversies.} Munson and Resnick \cite{MunsonR10} focus on blog posts to study if people seek diverse information, while  \cite{BalasubramanyanCPR12} use an LDA-based methodology to predict how different communities respond to political discourse. In \cite{FangSSY12}, Fang et al. propose a model to mine contrastive opinions for political issues, and many research groups devise methods for polarity detection and political leaning classification ~\cite{MaloufM08,YuKD08,BrandesKLR09,TsytsarauPD10,ZhouRM11,HodaD12,AwadallahRW12}. 

\textbf{Filter Bubble.} Pariser~\cite{Pariser11} points out the existence of the filter bubble, which he defines as ``this unique, personal universe of information created just for you by an array of personalizing filters'', and many works propose ways to mitigate its effects~\cite{GarrettR11,ResnickGKMS13,MunsonLR13}. The filter bubble effect is related to the selective exposure theory that sociologists developed much earlier~\cite{SEARSF67,Frey86}. 
Yom-Tov et al.~\cite{YomTovDG13} focus on news outlet sites that people visit, quantify the filter bubble and study whether users browse webpages supporting disagreeable information when opposing views are introduced in their search results. 

\textbf{Temporal Evolution of Knowledge.} This topic has been studied based on web logs in terms of changes in vocabulary, sites visited, and search strategies \cite{WhiteDT09,KotovBWDT11,Alonso13,LiuBZY13}. Related work also includes study of the evolution of fame and public interest in people~\cite{CookSFT12} as they are inferred from news articles.
\pnb{ 
 The related work is still choppy. }

\textbf{This Work.}
In contrast to most previous work, which considers 
primarily news outlets and blogs, and studies whether people access sources of different political categories to get informed~\cite{MunsonLR13,YomTovDG13}, we put a particular topic under the microscope and study how {\it that} affects the browsing behavior of the users. 
Another major difference from prior efforts is that we separate the political orientation of the users from their orientation to the gun debate.
For example, a Gallup poll in 2005 \cite{gallup05} indicated that 23\%/27\%/41\% of, respectively, Democrats/Independents/Republicans own a gun for an overall average of 30\% of US adults.  Thus, while gun ownership correlates with political leanings, there is significant ownership in each population. Given that, it is quite likely that views toward the gun debate may differ from party affiliation as well.
Thus, we do not engage in the common practice of characterizing websites as liberal and non-liberal. Rather, we define our own {\em content}-oriented labels (Appendix~\ref{app:categories}). 
Finally, although our work is motivated by the findings of prior studies on the existence of the filter bubble, our focus is not limited to corroborating or opposing this view. Our goal is to understand the types of webpages people visit, as well as how they transition among 
 content expressing different viewpoints.

We contribute an analysis of the temporal evolution of the users' {\it browsing} behaviors, and especially the {\it influence of specific external events with nationwide impact} on the shaping of the users' stances and their overall polarity.  We also
analyze the transitions of the users among webpages of different viewpoints.


\section{The {\large\nonNews} Data}
\label{sec:gun_data}
  
\hide{29,046,542}

\hide{\reminder{think of nice pictures that show the extraction process}}

\subsection{Data Extraction}
\label{sec:dataExtraction}

We use a proprietary dataset comprising anonymized web browser toolbar logs from November and December 2012
for more than 29 million users in the US-English market. 
The web logs include search and browsing behavior for the users, covering issued queries and visited, non-encrypted URLs. The selected users constitute a fair and broad sample of the US population. We consider primarily two time periods: before and after the Sandy Hook Elementary School shooting on December 14th. We note that we consider logs from a longer period of time before the event to develop a more robust estimate of users' habitual activity -- a similar quantity of activity is observed in the period after the shooting because information seeking is more frequent after the event (Fig.~\ref{fig:crownJewel}). 
For the purposes of our study, we consider the URLs that are \emph{on-topic}, i.e., websites that discuss gun control/rights issues. Hence, our first goal is to extract the relevant data with techniques that can be re-used in a programmatic manner for the analysis of other topics.

A na\"{\i}ve approach to obtaining a corpus of on-topic data is to consider all webpages containing the word ``gun''. Such an approach leads to numerous false positives, including websites about toys, video games, glue guns etc. 
We took an alternate approach that led to a corpus with many fewer false positives. The extraction process focused on identifying on-topic seed queries with high precision and then expanding these to related URLs and queries to obtain high coverage of all the on-topic activity. Specifically, our multi-step approach, as illustrated in Fig.~\ref{fig:data_extraction}, is to do the following:

%
\begin{table}[b]
\centering
  \caption{ 
  The most popular seed queries (col. 1), and relevant queries before and after the \SH shootings (col.~2 and 3).}
  \resizebox{\columnwidth}{!}{
\begin{tabular}{l|l|l}
    \toprule
    \multicolumn{1}{c|}{\textbf{Top 15 seed queries}} & \multicolumn{1}{c|}{\textbf{Top 15 relevant queries}} & \multicolumn{1}{c}{\textbf{Top 15 relevant queries}} \\
    \multicolumn{1}{c|}{} & \multicolumn{1}{c|}{\textbf{before \SH}} & \multicolumn{1}{c}{\textbf{after \SH}} \\
    \midrule
    Bob Costas gun control & Bob Costas gun control & Connecticut shooting \\
    gun control petition & shooting & school shooting in Connecticut \\
    Rupert Murdoch gun control & 2nd amendment & school shooting \\
    Piers Morgan gun control & gun control & Connecticut school shooting \\
    gun control & nutnfancy & shooting in Connecticut \\
    Feinstein gun control & Oregon shooting & elementary school shooting \\
    gun control debate & second amendment & gun control petition \\
    Rahm Emanuel gun control & concealed carry & Rupert Murdoch gun control \\
    Murdoch gun control & National Rifle Association & Sandy Hook shooting \\
    gun control laws & Obama gun control & shooting \\
    \bottomrule
    \end{tabular}%
  \label{tab:queries}%
  }
\end{table}%
\begin{itemize}
\item \textbf{Step 1. Identification of Relevant Queries:} 
We start with easy to identify relevant queries through keyword matching, and automatically expand them to as many relevant queries as possible by exploiting usage data.

 \begin{figure}[t!]
    \centering
    \includegraphics[width=0.95\columnwidth]{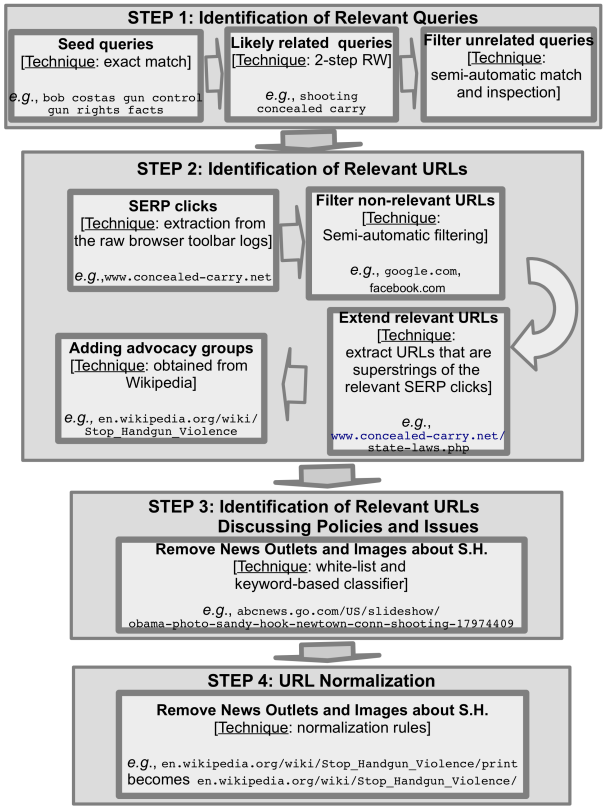}
    \caption{Illustration of the data extraction process.}
    \label{fig:data_extraction}
  \end{figure}

\textbf{1A. Seed Queries.} First, we identify seed queries by extracting those queries that contain the phrases ``gun control'' or ``gun rights'', but that are not related to electronic games. By doing this, we automatically filter out the queries that have an exact match with ``xbox'', ``wii'', ``gun controller'', ``game'', or ``playstation''. The resulting set consists of 6,878 queries, the ten most popular of which are given in Table~\ref{tab:queries} (col.~1). 

\textbf{1B. Identifying Likely Related Queries.} The second step consists of expanding the set of seed queries to relevant queries (misspelled, different expressions of the same intent, {\em etc.}). For this purpose, we create the query-click graph, a bipartite graph, where each query in the web logs is connected to the impression URLs that some user clicked; queries linked via a clicked URL are referred to as co-clicked.  
Starting from the seed set, we perform a two-step random walk \cite{craswell-szummer:sigir-2007}, and expand the seed set to all the similar co-clicked queries, as evaluated by their character-trigram cosine similarity with the seed queries.
%
 The threshold for similarity is set to 0.5 to require relatively high similarity. 
Intuitively, the new queries are connected to the same URLs as the seed queries and have textual overlap. Thus, they are likely on-topic, and probably represent alternative ways of querying for the same web search results.

\textbf{1C. Filtering Non-Relevant Queries.} Finally, from the likely relevant set of queries, we filter out the most common overall and seasonal queries, such as the navigational queries, {\it google} and {\it facebook}. Moreover, by manually inspecting the queries without the word ``gun'', we remove those queries that are not directly related to gun control, and lead to retrieval of numerous URLs unrelated to gun control (high recall/low precision) -- e.g., {\it what do democrats and republicans stand for}, 
{\it conservative viewpoint}. 
The final, extended set, to which we will refer as set of {\it relevant} queries, consists of 7,778 queries. The most popular queries before and after \SH are given in Table~\ref{tab:queries} (col.~2 and 3). 

\item \textbf{Step 2. Identification of Relevant URLs:} 
Users reach URLs through many ways ({\emph e.g.,} browsing, search, bookmarks).  Our objective is to use the resulting on-topic \emph{queries} to identify sessions of information-seeking behavior, which according to IR research tend to be topically coherent.
%
%
Again, a na\"{\i}ve approach would be to extract any clicked URL from the search engine result page (SERP) of a topical query, as well as the pages browsed subsequently by consecutive clicks (click trail). 
However, users may click on ads and other contextual links (some of which may be topically relevant, but often not), and browse from a topical article to a non-topical one as they drift to a different topic.  Therefore, similar to identifying relevant queries, we developed a semi-automated way of expanding to a broad topically relevant set without incorporating significant amounts of off-topic search and browsing.  In particular, we implemented the following procedure for identifying relevant pages:




\textbf{2A. SERP Clicks.} Starting from the relevant queries of the previous step, we obtain only the URLs users clicked directly from a topically relevant query's SERP.  

\textbf{2B. Filtering Non-Relevant URLs.} Then, we semi-automatically filter popular URLs that are off-topic, as well as YouTube videos. Although media analysis is interesting, we focus primarily on non-video web pages ({\em i.e.}, mainly text).  We refer to this set of filtered URLs on guns, gun control and gun rights as {\it seed} URLs.


\textbf{2C. Extend Relevant URLs.} We continue by extending the set of seed URLs to include more webpages that might not belong to the SERP clicks of relevant queries.  To this end, we consider relevant all URLs 
that are superstrings of the seed URLs. The intuition is that those were either reached from or led to a seed URL, and have high overlap in the site organization -- implying a topical relationship. Moreover, this process leads to higher recall, as it also extracts URLs entered in the toolbar, or saved as bookmarks. 

\textbf{2D. Adding Advocacy Groups.} The method described above is {\it not} guaranteed to extract {\it all} the URLs that are relevant to gun control and rights, but it attempts to extract as many, highly related websites as possible, while maintaining neutral criteria with respect to the topic of study. Extracting all the webpages that are on-topic is challenging and could stand as a research problem on its own. We seek to make sure that we capture visits to webpages for the most prevalent gun control and rights advocacy groups.  Thus, we take compiled lists from Wikipedia\footnote{Gun control / rights advocacy groups in the United States: \scriptsize{ http://en.wikipedia.org/wiki/Category:Gun\_control\_advocacy\_groups\_in\_the\_United\_States} and \scriptsize{ http://en.wikipedia.org/wiki/Category:Gun\_rights\_advocacy\_groups\_in\_the\_United\_States}}, and explicitly extract user visits to both the advocacy group websites and their Wikipedia pages.


\item \textbf{Step 3. Identification of Relevant URLs Discussing Policies and Issues:} By using a white-list and keyword-based classifier, we filter out news outlets, which merely report news about the incident without discussing gun-related issues and policies. The reasoning behind this is two-fold: (a) We seek to explore how users access information in reaction to a news event, instead of how they get informed about the details of the event; (b) Although one can argue that some news sites are representative of specific ideological views, we do not rely on the latter, because often the political orientation differs from the orientation to the gun control issue \cite{gallup05}. For the same reason, we filter out images from the shootings (but keep images with sarcastic comments on the issues).

\item \textbf{Step 4. URL Normalization:} Finally, 
we normalize the URLs so that different webpages with the same content, mobile versions of the websites, print requests of a page, user id encoding pages, {\em etc.}\ are considered the same. 

\end{itemize}

The resulting dataset, \nonNews, consists of records $<$user-id, session-id, URL, timestamp$>$ (Table~\ref{tab:nonNewsStats}). In the following sections, we refer to the intersection between the sets of users before and after the shootings as \textbf{common users}.  Studying them 
 enables us to directly compare changes in user behavior by controlling for the set of users.




\begin{table}[t]
  \centering
  \caption{\nonNews dataset. The last column holds the number of common users, URLs and domains between the two time periods.}
  \label{tab:nonNewsStats}%
  \resizebox{\columnwidth}{!}{
    \begin{tabular}{rrrrr}
    \toprule
    & \multicolumn{1}{c}{\textbf{Before \SHshort}} & \multicolumn{1}{c}{\textbf{After \SHshort}} & \multicolumn{1}{c}{\textbf{Total}} & \multicolumn{1}{c}{\textbf{Overlap}} \\
    \midrule
   \textbf{Users} & 10,336 & 54,849 & 61,276 & 3,909 \\
       \textbf{Sessions} & 20,656 & 70,138 & 90,794 & N/A \\
       \textbf{Unique URLs} & 5,955 & 20,090 & 24,439 & 1,606 \\
       \textbf{Unique Domains} & 277 & 501 & 613 & 165 \\
       \textbf{Total Visits} & 118,749 & 177,975 & 296,724 & N/A \\
    \bottomrule
    \end{tabular}%
    }
\end{table}%

\hide{
\begin{table}[htbp]
  \centering
  {\small
  \caption{\news dataset. \danai{Remove?} The last column holds the number of common users, URLs and domains between the two time periods.}
  \resizebox{\columnwidth}{!}{
    \begin{tabular}{rrrrr}
    \toprule
    & \multicolumn{1}{c}{\textbf{Before \SHshort}} & \multicolumn{1}{c}{\textbf{After \SHshort}} & \multicolumn{1}{c}{\textbf{Total}} & \multicolumn{1}{c}{\textbf{Overlap}} \\
    \midrule
   \textbf{Users} & 56,232 & 678,078 & 708,456 & 25,854 \\
       \textbf{Sessions} & 61,424 & 1,007,485 & 1,068,909 & N/A \\
       \textbf{Unique URLs} & 302   & 9,273 & 9,559 & 16 \\
       \textbf{Unique Domains} & 61    & 296   & 308   & 49 \\
       \textbf{Total Entries} & 68,685 & 1,529,944 & 1,598,629 & N/A \\
    \bottomrule
    \end{tabular}%
    }
  \label{tab:newsStats}%
  }
\end{table}%
}

\subsection{Data Annotation}
\label{sec:annotation}

Answering the questions of interest is not possible unless the webpages are labeled based on their stance on gun control/rights.  Rather than focusing on alignment with a political party, we focus on the disposition of the content itself.  Visits to a site that is predominantly affiliated with one party ({\em e.g.}, Democratic/Republican) or a particular pundit, does not by itself imply a lack of diversity in content; sites may contain content discussing a broad range of material.  Considering the content also enables us to measure that extent to which sites provide information representing diverse views.

Manually labeling {\it all} the webpages is difficult. Our attempts to automate the labeling process by building {\em content}-centric classifiers failed to achieve high accuracy, revealing the challenges of classifying controversial pages by their stance. We could not apply the extensive work on detecting and labeling controversial topics \cite{BrandesKLR09,ZhouRM11,HodaD12}, 
 as our setting is different: we seek to characterize the presented \emph{viewpoints} in documents on a \emph{given controversial topic}. 
 To overcome these challenges, we judged all webpages that had at least two unique visitors and sampled from the remaining webpages.  
 The on-topic and accessible pages were initially judged by their content and classified into three categories (high-level labels) -- balanced, gun control, gun rights --, and then into expanded categories that reflect the stance of the webpages at a finer granularity (expanded labels): purely factual and highly balanced, extreme and moderate gun control, and extreme and moderate gun rights. Details about the labels are provided in the Appendix.
 


Two assessors were provided with a subset of the over 1,500 selected webpages, and were asked to classify them. 
One assessor self-identified as ``moderate gun rights'', while the other self-identified as ``moderate gun control''. The inter-rater agreement \cite{Randolph05}, which already accounts for the chance-expected proportion of agreement between the assessors, is 75.00\% for the high-level classification, and 58.86\% for the expanded labels that reflect further key category distinctions. 
We note that these inter-rater agreements are high, since the chance-expected agreement using the marginal distribution is 36.8\% \danai{or 20\% if we use a uniform model} for the high-level labels, and 20.48\% \danai{or 12.5\% if we use a uniform model} for the expanded labels. 
\danai{For the chance agreement I used the probs of labels in the overlap set, and multiplied the probs of same labels and added them. For the uniform case, I assume that the prob of each label is 1/5 or 1/8 resp. }

\begin{table}[t]
  \centering
  {\small
  \caption{Inter-rater agreement for the high-level labels (col.~1), and the expanded set of labels (col.~2). Overall agreement is simply the percent of labels on which the raters agree.}
    \begin{tabular}{lrr}
    \toprule
    \textbf{} & \multicolumn{2}{c}{\textbf{Labels}} \\
    \midrule
    \textbf{} & \multicolumn{1}{c}{\textbf{High-level}} & \multicolumn{1}{c}{\textbf{Expanded}} \\
    \textbf{Overall agreement} & 80.00\% & 64.00\% \\
    \textbf{free-marginal $\kappa$} & 75.00\% & 58.86\% \\
    \textbf{fixed-marginal $\kappa$} & 68.27\% & 53.99\%  \\ 
    \textbf{chance-expected agreement} & 36.80 \% & 20.48\% \\
    \bottomrule
    \end{tabular}%
    \label{tab:inter-rater}%
    }
\end{table}%

From the extracted on-topic web pages 
25\% are purely factual, 10\% highly balanced, 31\% and 21\% moderate and extreme gun rights resp., and 9\% and 4\% moderate and extreme gun control.  
 
 \hide{
 \begin{figure}[ht!]
     \centering
       \includegraphics[width=0.56\columnwidth]{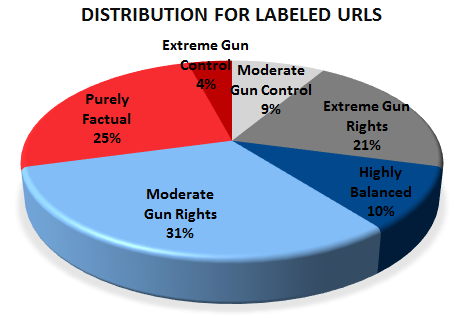}
          \caption{Distribution of labels after manual labeling by 2 assessors.}
                   \label{fig:distLabels}
   \end{figure}
}

\hide{   
   \begin{figure}[ht!]
       \centering
        \includegraphics[width=\columnwidth]{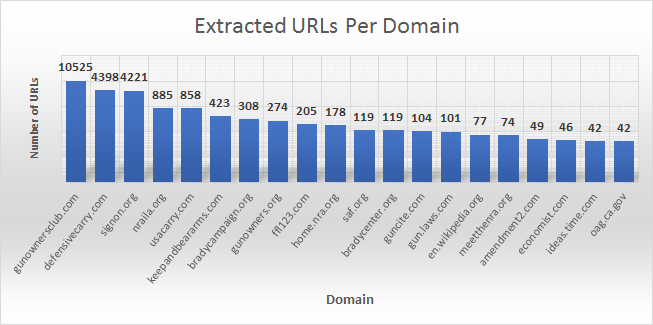}
        \caption{Number of URLs for the top-20 domains.}
        \label{fig:urlsPerDomain}
      \end{figure}
}
   

 \hide{(a) some of the webpages are forums which have many different threads (e.g., gunownersclub.com with 10,525 URLs, as shown in Fig.~\ref{fig:urlsPerDomain}), (b) some URL links contain information about the user browsing the website, and appear to be different, but they are actually the same (e.g., signon.org which append the user ID to the URL). Especially in the first case, the users often browse multiple pages of the forums; as a result, transitions between webpages sharing domain are dominant, but these transitions are not the focus of this work. We normalize the URLs so that different pages of the same article, mobile versions of the same articles, print requests of a page etc. are considered the same.
  We handle each case separately: 
 \begin{itemize*}
 \item {\bf Forums.} We replace the URLs belonging to forums with the link of their main page, and classify the latter based on their political leaning (which coincides with the dominant class of the manual labeling).
 \item {\bf Advocacy groups.} We treat the gun control and rights advocacy groups in the same way.
 \item {\bf signon.org.} This is a petition on gun control, which encodes user information in the link. Since it is just one webpage, we collapse all the pages to one. 
 \item {\bf Domains.} For the domains that contain webpages in various categories, we assign the dominant category. First we identify the broad group (control, rights, balanced) containing the majority of webpages, and then, within this group, we assign the stance (extreme, moderate) that is most commonplace. If there is a tie between the possible categories, we do not classify the domain, and keep the initial URLs and their labels for our analysis.
 \item We remove images from the shooting - off-topic - kept only images with sarcastic comments on gun control/rights.
 \end{itemize*}
 
 By following the rules described above, we end up with the final labeling of the domains and remaining URLs whose stance could not be summarized succinctly by a single label. The distribution of the final labels is given in Fig.~\ref{fig:labelDistributionCollapsedDomains}.
 
  \begin{figure}[ht!]
     \centering
      \includegraphics[width=0.56\columnwidth]{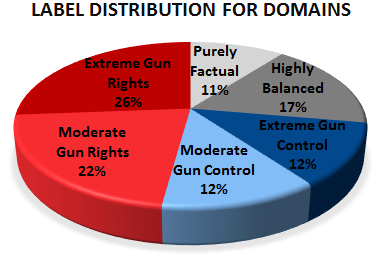}
      \caption{After representing the URLs by their domain: Distribution of final labels. }
      \label{fig:labelDistributionCollapsedDomains}
    \end{figure} 
  }
 
 \hide{
 
 \paragraph{Label Propagation}
 Despite manually labeling more than a thousand websites based on their stance on the gun control/rights topic, we are left with almost 22,000 less popular URLs. As mentioned in \cite{BarKS07}, the web is full of \textsc{DUST}, i.e., Different URLs with Similar Text. We took advantage of this ``problem'', and first propagate the known labels to (similar) URLs with similar content. Then, in order to obtain even more labeled data, we label websites for the stance of which we have strong indication, by applying a {\it domain-based} label propagation procedure. Specifically, we cluster the labeled websites by domain, and compute the percentage of websites in each category (see Fig.~\ref{fig:distLabelsPerDomain}). For the domains with at least 5 labeled URLs, we assume that the unlabeled URLs belong to the dominant category. For example, we classify the ~4.2K URLs with domain {\it signon.org} as ``Moderate gun control'', while the URLs with domain {\it defensivecarry.com} as ``Moderate gun rights''.
 
 After dealing with the \textsc{DUST}, and performing domain-based label propagation, we end up with almost 24,000 \hide{23,771} labeled URLs. The label distribution for the final set of URLs is given in Fig.~\ref{fig:distLabelsExtendedLeap}.
 }

 \hide{
 \begin{figure}[ht!]
   \centering
      \includegraphics[width=\columnwidth]{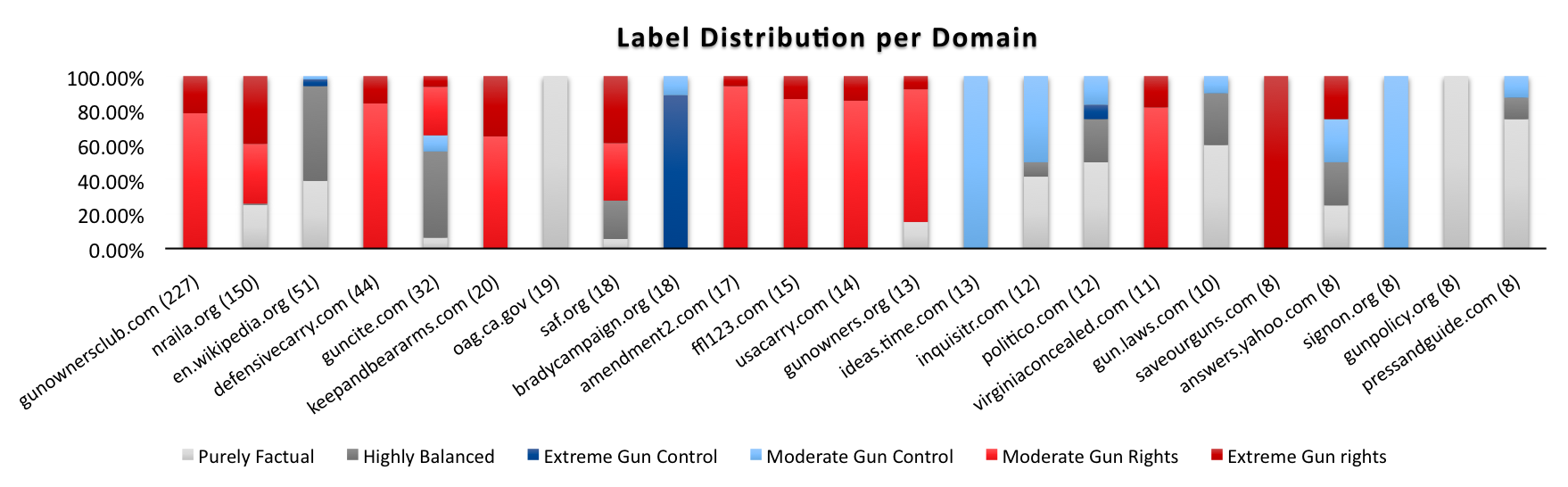}
   
       \caption{Label distribution per domains. The domains are ordered in decreasing number of manually characterized URLs (in parentheses).}
              \label{fig:distLabelsPerDomain}
 \end{figure}
 }

\hide{{\bf Labeling the \news dataset.} \reminder{I want to check if I can use Elad's rankings, but I think that there is not much overlap between the domains he had, and the news domains we have extracted. On the other hand, I can label the overlap and proceed with that - accepting the fact that this is how it is with real data! - 170 of the 3}
 }  

\section{Diversity of Domains}
\label{sec:domainDiversity}

\hide{
 \begin{figure}[ht!]
               \centering
               \includegraphics[width=\columnwidth]{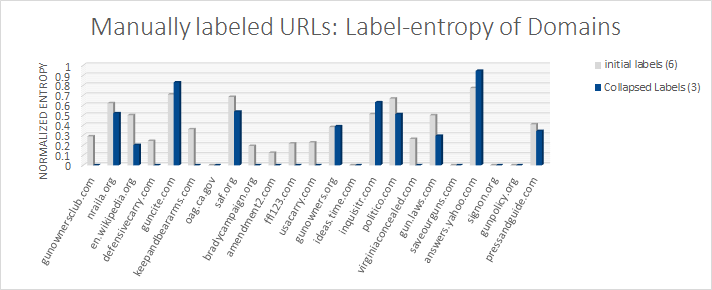}
               
                 \caption{\nonNews: Diversity of domains in terms of label entropy (for the manually labeled URLs). }
                       \label{fig:entropyLabelsPerDomain}
             \end{figure}
 }
 
 
 Our first study focuses on the types of available information, and particularly the diversity of web domains. We identify domains with at least eight labeled webpages and give their label distribution in Fig.~\ref{fig:distLabelsPerDomain}.
 %
 %
 \begin{figure}[b!]
    \centering
    \includegraphics[width=1.03\columnwidth]{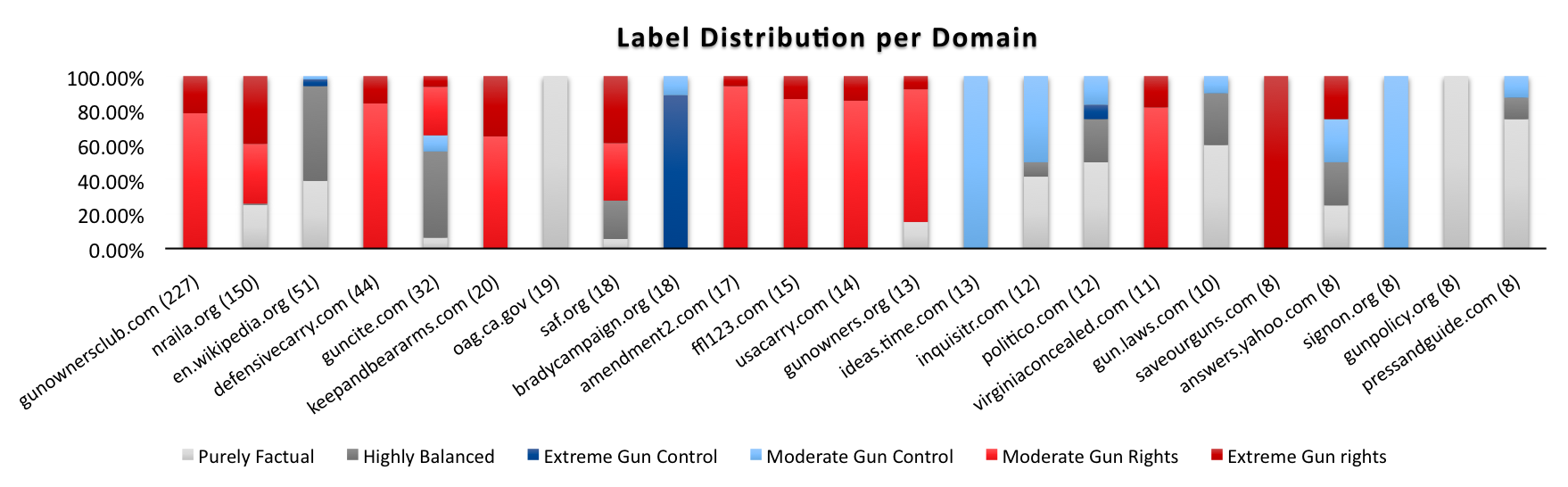}
    \caption{Label distribution per domain. The domains are in decreasing order of manually characterized URLs (in parentheses).}
    \label{fig:distLabelsPerDomain}
  \end{figure}
 %
 It is evident that most of the web domains are one-sided, with almost all their webpages expressing similar opinions (e.g., supporting only gun rights). An exception to this finding is
  that user-generated content, such as that found on {\small \tt wikipedia.org} and {\small \tt answers.yahoo.com}, tends to be diverse. \pnb{either balanced or diverse, respectively}
 
 To quantify the heterogeneity of the available information per domain in a principled way, we use Shannon's entropy~\cite{Shannon48}, an information-theoretic measure of the uncertainty for a random variable. The higher the entropy associated with a random variable, the higher the uncertainty about its value, or, equivalently, the more diverse 
 it is. 
 Formally, for each domain $d$ with entropy \hfill \\
\resizebox{.9\hsize}{!}{$H(X_d) = E[-\lg{P(X_d)}]=-\sum_i{P(X_d=x_i)*\lg{P(X_d=x_i)}},$}\\
 we compute the {\it normalized entropy} for its webpage labels: 
	$$H^{norm}(X_d) = H_{t}(X_d)/H_{t}(X'_d | X'_d\sim \mathcal{U}),$$
	where 
  $X_d, X'_d$  are the labels of the webpages with domain $d$, 
  $X'_d$ is uniformly distributed, and $\lg$ is the base-2 logarithm. We note that $H_{t}(X'_d | X'_d\sim \mathcal{U})$ corresponds to the maximum entropy where the labels occur with equal probability.
  
  We compute the {\it normalized} entropy for the labels of the URLs instead of the entropy for two reasons: (1) the normalized entropy handles comparisons across different event space sizes, which is needed when comparing high-level and expanded labels and (2) the normalized entropy ensures that comparisons between domains with different numbers of observations are at the same basis. 
  Normalizing the measure helps to handle estimation error, as the entropy can have high variance when there are only a few observations.
 
  \begin{figure}[t!]
      \centering
      \includegraphics[width=\columnwidth]{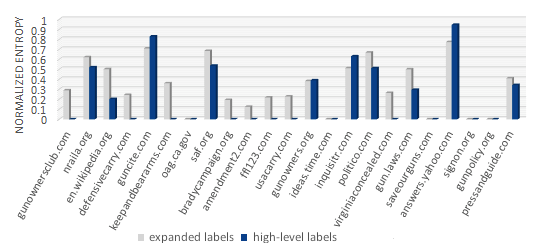}
      \caption{\nonNews: Diversity of domains in terms of label entropy (for the manually labeled URLs). }
      \label{fig:entropyLabelsPerDomain}
 \end{figure}

Figure~\ref{fig:entropyLabelsPerDomain} depicts the normalized entropy in the labels of the webpages per domain. For each domain, the left and right bars correspond to the normalized label entropy for the expanded, and the high-level labels respectively.
%
Overall, for the high-level labels, the normalized entropy is 0 (no diversity) for 54\% of the domains, and smaller than or equal to 0.5 for 63\% of the domains. The median normalized entropy is 0, and the mean 0.27. 
%
Similarly, for the expanded set of labels, 34\% of the domains have entropy 0, and 73\% have normalized entropy smaller than or equal to 0.5. The median and mean normalized entropy are 0.36 and 0.30 respectively.
The main finding is that the domains 
offer to the users mostly 
a single myopic view on gun control issues. 
%
%
 %
  Based on this observation, we were able to automatically label the remaining $\sim$23K webpages that were not labeled manually by the assessors, and obtain an annotated dataset that can serve the purposes of our next analyses. We apply a label propagation approach from the webpages to their domains: 
 \begin{itemize*}
 \item {\bf Forums.} We replace URLs that belong to a forum with its main page, and classify the latter based on the overall stance of its labeled webpages, ({\em i.e.}, the dominant category of the manual labeling).
 \item {\bf Advocacy groups.} We label each domain based on the identified stance using Wikipedia's characterization.
 \item {\bf Domains.} For the domains with normalized entropy smaller than 0.5, we first assign the dominant high-level category, and then the stance (moderate, extreme) of the majority of the labels. 
 If we have a tie among the possible categories, we do not classify the domain, and keep the initial URLs and their labels for our analysis.
 \end{itemize*}
 
 By following these rules, we obtain the final labeling of the {\it domains}, as well as the remaining URLs whose domain's stance could not be summarized succinctly by a single label. The distribution of the final labels is: 11\% purely factual, 17\% highly balanced, 22\% and 26\% moderate and extreme gun rights respectively, and 12\% of both moderate and extreme gun control.

 \hide{
  given in Fig.~\ref{fig:labelDistributionCollapsedDomains}.
 
  \begin{figure}[ht!]
     \centering
      \includegraphics[width=0.56\columnwidth]{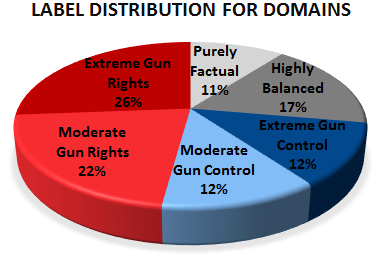}
      \caption{After representing the URLs by their domain: Distribution of final labels. }
      \label{fig:labelDistributionCollapsedDomains}
    \end{figure} 
  }



\section{Within-User Diversity}
\label{sec:userDiversity}
Our second study focuses on the diversity of information consumed by each user browsing controversial topics, and how the diversity in the information sought is influenced by a shocking news event.
 The within-user diversity can be expressed in terms of the number of different domains that a user browses, as well as the number of different {\it categories} (e.g., gun control, balanced webpages) of pages that she visits. As in Sec. \ref{sec:domainDiversity}, we use Shannon's entropy to quantify the diversity in the categories of webpages that each user visits.

\subsection{Examining the Existing Theories}
We find two contradictory theories in the prior literature, which we consider regarding the implications of using entropy to capture variance:
\begin{itemize*}
\item {\bf Theory 1.} ``People use the web to access a variety of different sources, and become more aware of political news and events'' \cite{Stromer-Galley03,HorriganGRP04}.\\
{\bf Implication 1.} If this assertion is true, we would expect users to visit domains with different labels regarding perspective, and that the associated normalized label entropy of the domains that the users browse would be high. In the case where users visit only a few domains, we would expect the label entropy of the domains to be high.
\item {\bf Theory 2.} ``People use the web to access mostly agreeable political information'' \cite{AdamicG05,GilbertTK09,Garrett09}.\\
{\bf Implication 2.} If this assessment is true, we would expect most users to access mostly domains supporting one side of the topic.  Thus, the label entropy of the domains that the users access should be low.
\end{itemize*} 

We now analyze the diversity of information that users consume to explore the two assertions. We first focus on all users who visited at least three relevant domains during November and December. 
From all the users only 5\% fall into this category; the vast majority of them, 61\%, accessed exactly 3 domains, 23\% browsed 4 domains, and 9\% visited 5 domains. This observation, in combination with the low diversity of domains (Sec.~\ref{sec:domainDiversity}), provide evidence that Theory 1 is unlikely. Most users appear too ``narrow-minded'' as far as the number of web domains is concerned, and the domains themselves are mostly one-sided.



To evaluate Theory 2, we need to examine the diversity of each user's consumed information by computing the entropy in the labels of the domains accessed. The intuition behind this need is that the number of domains does not provide enough information about the diversity of a user's exposure, as it does not fully characterize the type of information consumed\hide{capture the type of websites that were browsed}. Two extreme cases would be a user who visited three websites supporting gun control, and another user who visited a website of each category: gun control, gun rights and balanced. Clearly, the second user's information stream is more diverse. Thus, for each user $u$ who visited at least three different relevant domains during November and December, we compute a normalized label entropy 
\begin{equation}
H^{norm}(X_u) = H(X_u)/H(X'_u | X'_u\sim \mathcal{U})
\label{eq:norm_label}
\end{equation}
 where $X_u, X'_u$  are the labels of the domains visited by user $u$ (thus, $X_u, X'_u$ take values in $\{\mbox{gun control}, \mbox{gun rights}, \mbox{balanced/factual}\}$), and $X'_u$ is uniformly distributed rendering $H(X'_u | X'_u\sim \mathcal{U})$ the maximum entropy. 
 
The average normalized entropy is 0.49, and the median is 0.58. 
Hence, the majority of users neither access webpages of a single stance ($H^{norm}$ should be 0), nor websites of all possible labels ($H^{norm}$ should be 1). All in all, we see more evidence for Theory 2 than for Theory 1. 

%

\subsection{Event and Change in Within-User Diversity}

\begin{figure}[t]
    \centering
     \includegraphics[width=\columnwidth]{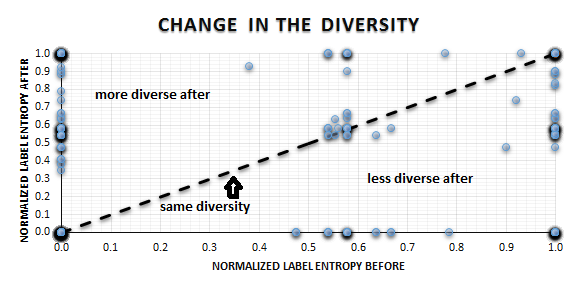}
     \caption{ 
     Change in the user diversity after \SHshort for users who visited at least two different domains both before and after the event. Every point in the plot corresponds to a user.}
     \label{fig:change_allUsers_3domains}
\end{figure}

Despite the fact that the diversity of information consumed by each user is low overall, we seek to understand if an event with nationwide impact influences {\em a change} in the diversity of the information that users access. For a fair comparison of accesses before and after the shocking news event, we need to control for the users. Thus, we focus on the  
 3,900 common users, {\emph i.e.,} those who were active {\it both before and after} \SH (Table~\ref{tab:nonNewsStats}), as these are the only users whom we can characterize before and after the event.
%
  
For each user $u$ who visited at least two domains both before and after \SH, we
compute per time period the normalized label entropy, $H^{norm}(X_u)$, as defined in Eq.~\eqref{eq:norm_label}.
%
 Then, to quantify the change in the within-user diversity, we compute the difference between the normalized entropies
 $$H_{AFTER}^{norm}(X_u) - H_{BEFORE}^{norm}(X_u).$$
  
 Next, we consider two cases for the data: (1) {\it all} the websites, and (2) all the webpages {\it except for} Feinstein's page about the new gun control legislation (gray spike on Dec. 27 in Fig.~\ref{fig:crownJewel}), which was heavily visited the day of its creation. 
 
 \noindent \textbf{With Feinstein's website.} Figure~\ref{fig:change_allUsers_3domains} shows the change in the user diversity after the Sandy Hook shootings. On average, the normalized entropy increased by 8.14\% after \SH. However, the results vary across users: the entropy remained the same for 36\% of the users (and a majority of those, about 70\%, are highly polarized before and after the event, visiting domains in one category only); for 36.2\% of the users the diversity increased by 66.53\%; and for 27.8\% decreased by 57.3\%.

To understand what drove so many users to this site, we investigated more deeply.  On Dec. 27, 2012, the Drudge Report, a primarily conservative link aggregation site, featured a gun rights reaction grabbing  headline ``Senate to Go For Handguns''\footnote{See
{\tt http://www.drudgereportarchives.com/data/ 2012/12/27/20121227\_160126.htm} for an archive.}. Additionally, a summary underneath mentioned key sensitive points for gun rights supporters (gun registries and owner fingerprinting).
84\% of the users who reached the Feinstein list came via the Drudge Report front page -- nearly all of them had primarily consumed gun rights information prior to this.   The implication  is that when users consume information outside of their typical viewpoint the link comes from {\em within} their own sphere.  Moreover, it suggests future research to redesign contextual link recommendation -- which displays related-content to a page -- may have a greater potential for changing the diversity of information a user browses than the composition of search results.

\noindent \textbf{Without Feinstein's website.} 
By repeating the analysis described above after removing the website which was heavily visited when the gun ban list was announced, we find that, on average, the normalized entropy of the users decreased by only 0.59\%. For 43.75\% of the users, the diversity remained the same; for 27.6\% of the users, the entropy increased by 62\%, and for 28.6\% it decreased by the same percent.

The change in user diversity, both with and without the outlier webpage, suggests that people peek outside of their bubble when events have potential for individual impact on the user -- such as the proposed legislation on banning some types of guns --, but remain in an ``echo chamber'' otherwise.

%






\section{Web Transitions}
\label{sec:change_in_transitions}
 Our last study addresses the way users navigate polarizing topics. Specifically, 
 how does the current website's stance affect how people transition to the next webpage on polarizing political topics? What webpages are they more likely to visit after browsing a site supporting extreme gun control or rights? 
 Although most of the users are not very diverse in terms of the label entropy of the domains they visit (Sec.~\ref{sec:userDiversity}), many of them {\it do} transition between pages supporting opposing views. We seek to understand the most common transitions, as well as possible changes in the transitions due to the news on the Newtown shootings. 
By focusing on the influences of the stance of the current page on transitions to next pages, we obtain a micro-level view of information consumption patterns as opposed to the user-level view presented in Section~\ref{sec:userDiversity}.


\begin{table}[t]
   \centering
  {\scriptsize
   \caption{Markov chain states for browsing history: Abbreviations for the high-level and expanded labels.} 
      \label{tab:abbreviations}%
     \begin{tabular}{ll||ll}
     \toprule
     \multicolumn{2}{c||}{\textbf{High-level States (3)}} & \multicolumn{2}{c}{\textbf{Expanded States (6)}} \\
     \midrule
     \multirow{2}{*}{\bf GC} & \multirow{2}{*}{Gun Control} & {\bf EC} & Extreme Gun Control \\
      & & {\bf MC} & Moderate Gun Control \\ \hline
     \multirow{2}{*}{\bf BF} & \multirow{2}{*}{Balanced/Factual} & {\bf HB} & Highly Balanced   \\
     &  & {\bf PF} & Purely Factual  \\ \hline
      \multirow{2}{*}{\bf GR} & \multirow{2}{*}{Gun Rights} & {\bf MR} & Moderate Gun Rights \\
     & &  {\bf ER} & Extreme Gun Rights  \\
     \bottomrule
     \end{tabular}%
     }
 \end{table}%

For each user 
we represent her browsing history as a Markov chain with either the high-level or expanded labels as states $X_i$ (Table~\ref{tab:abbreviations}). 
Then, we describe the distribution of the transition probabilities by an n-state transition matrix $\Pmat^n$, with elements $p_{ij} = Prob(X_{t+1}=j|X_t = i)$. We note that the row-wise sums are equal to 1, $\sum_jp_{ij}=1$.
%
%
To make sense of the underlying trends of this matrix, we employ mobility indices that have been widely used in economics and sociology (e.g., credit mobility \cite{JafryS04}, social status mobility \cite{BlauD67}). 


\subsection{Overall Transition Patterns}

 \begin{table}[b!]
    \centering
    {\scriptsize
    \caption{\label{tab:Transitions_ALL_2domains}{\it All} users: 6-state transition matrix $\Pmat^6$ for November-December.} 
      \begin{tabular}{lrrrrrr}
      \toprule
      & \multicolumn{1}{c}{\textbf{EC}} &
       \multicolumn{1}{c}{\textbf{MC}} & \multicolumn{1}{c}{\textbf{HB}} & \multicolumn{1}{c}{\textbf{PF}} &
       \multicolumn{1}{c}{\textbf{MR}} &
       \multicolumn{1}{c}{\textbf{ER}} \\
      \midrule
      {\bf EC} & 32.65\% & 3.21\% & 24.20\% & 7.58\% & 11.66\% & 20.70\% \\
      {\bf MC} & 15.22\% & 2.17\% & 27.17\% & 8.70\% & 22.83\% & 23.91\% \\
      {\bf HB} & 10.53\% & 2.26\% & 28.07\% & 14.04\% & 16.42\% & 28.70\% \\
      {\bf PF} & 5.83\% & 3.79\% & 27.11\% & 10.50\% & 20.99\% & 31.78\% \\
      {\bf MR} & 6.28\% & 3.07\% & 18.83\% & 11.09\% & 21.90\% & 38.83\% \\
      {\bf ER} & 5.49\% & 1.55\% & 15.93\% & 9.36\% & 19.18\% & 48.49\% \\
      \bottomrule
      \end{tabular}%
      }
  \end{table}%
  
  \begin{table}[b!]
     \centering
     {\scriptsize
     \caption{{\it All} users: 3-state transition matrix $\Pmat^3$ for November-December.} 
           \label{tab:Transitions_ALL_2domains_3states}%
       \begin{tabular}{lrrr}
       \toprule
       & \multicolumn{1}{c}{\textbf{GC}} &
        \multicolumn{1}{c}{\textbf{BF}} & 
        \multicolumn{1}{c}{\textbf{GR}}\\
       \midrule
       {\bf GC} & 28.42\% & 40.03\% & 31.55\%  \\
       {\bf BF} & 12.00\% & 45.64\% & 42.36\% \\
       {\bf GR} & 5.55\% & 27.74\% & 66.71\%  \\
       \bottomrule
       \end{tabular}%
       }
   \end{table}%
   
 First, we consider the transitions of \emph{all} users during November and December. There are $\sim 10K$ total transitions, corresponding to more than 7K \hide{7,070} users. We note that these are the users who visited at least two different domains, and, hence, we record for them at least one transition. The transitions are given in the form of a transition matrix in Tables~\ref{tab:Transitions_ALL_2domains} and ~\ref{tab:Transitions_ALL_2domains_3states}.
%

 We first employ the so-called {\it Summary Mobility Indices}, which describe the direction of the mobility:
\vspace{-0.2cm}
\begin{itemize*}
\item Immobility Ratio: $\textit{IR} = \sum_{i=1}^np_{ii}/n$
\item Moving Up: $\textit{MU} = \sum_{i<j}p_{ij}/n $
\item Moving Down: $\textit{MD} = \sum_{i>j}p_{ij}/n$,
\end{itemize*}
\vspace{-0.2cm}
where $n$ is the number of states. The immobility ratio represents the percent of same-state transitions (higher for more {\it im}mobility), while the other two indices give the percent of transitions from one extreme to the other, {\em i.e.}, the MU index captures the transitions from extreme gun control towards extreme gun rights, and the MD index describes the opposite directionality (higher for more mobility).

The Summary Mobility Indices for all users during November and December are: (a) for the high-level states $\textit{IR} = 0.4692,\; \textit{MU} = 0.3798$, and $\textit{MD} = 0.1510$, and (b) for the extended states $\textit{IR} = 0.2486, \; \textit{MU} = 0.4997$, and $\textit{MD} = 0.2518.$
Firstly, we observe that the overall system is characterized by mobility ($\textit{IR} \ll 1$). Specifically, for the extended states, about 25\% of the transitions are same state, and $50$\% of the transitions occur in the direction from extreme gun control towards extreme gun rights. From the transitions in the opposite direction, the most dominant transitions are towards the ``middle'' states: from factual to balanced webpages (27.11\%), from extreme to moderate gun rights (19.18\%), and from moderate gun rights to balanced pages (18.83\%). All in all, the users mainly browse domains of the same stance or transition from gun control and balanced pages to websites supporting gun rights.   


\subsection{\SH: Change in Transition Patterns}

\begin{table}[t!]
  \centering
  {\scriptsize
  \caption{{\it Common} users: Transition matrix $\Pmat^6_{\textit{before}}$ for the time period before \SH.} 
    \begin{tabular}{lrrrrrr}
    \toprule
    & \multicolumn{1}{c}{\textbf{EC}} &
     \multicolumn{1}{c}{\textbf{MC}} & \multicolumn{1}{c}{\textbf{HB}} & \multicolumn{1}{c}{\textbf{PF}} &
     \multicolumn{1}{c}{\textbf{MR}} &
     \multicolumn{1}{c}{\textbf{ER}} \\
    \midrule
    {\bf EC} & 49.38\% & 0.00\% & 18.52\% & 3.70\% & 16.05\% & 12.35\% \\
    {\bf MC} & 11.76\% & 0.00\% & 23.53\% & 0.00\% & 23.53\% & 41.18\% \\
    {\bf HB} & 8.82\% & 1.68\% & 26.47\% & 8.40\% & 27.31\% & 27.31\% \\
    {\bf PF} & 2.90\% & 1.45\% & 30.43\% & 4.35\% & 33.33\% & 27.54\% \\
    {\bf MR} & 2.97\% & 1.30\% & 9.65\% & 6.86\% & 42.67\% & 36.55\% \\
    {\bf ER} & 1.82\% & 0.13\% & 10.29\% & 2.86\% & 20.70\% & 64.19\% \\
    \bottomrule
    \end{tabular}%
      \label{tab:Transitions_COMMON_before}%
    }
\end{table}%

\begin{table}[t]
  \centering
  {\scriptsize
  \caption{{\it Common} users: Transition matrix $\Pmat^6_{\textit{after}}$ for the time period after \SH.} 
    \begin{tabular}{lrrrrrr}
    \toprule
    & \multicolumn{1}{c}{\textbf{EC}} &
     \multicolumn{1}{c}{\textbf{MC}} & \multicolumn{1}{c}{\textbf{HB}} & \multicolumn{1}{c}{\textbf{PF}} &
     \multicolumn{1}{c}{\textbf{MR}} &
     \multicolumn{1}{c}{\textbf{ER}} \\
    \midrule
    {\bf EC} & 22.03\% & 3.95\% & 26.55\% & 7.91\% & 12.99\% & 26.55\% \\
    {\bf MC} & 13.85\% & 0.00\% & 15.38\% & 13.85\% & 16.92\% & 40.00\% \\
    {\bf HB} & 9.01\% & 1.93\% & 21.89\% & 21.24\% & 15.02\% & 30.90\% \\
    {\bf PF} & 2.68\% & 3.68\% & 19.73\% & 8.36\% & 20.07\% & 45.48\% \\
    {\bf MR} & 3.97\% & 1.51\% & 15.69\% & 14.74\% & 18.15\% & 45.94\% \\
    {\bf ER} & 2.85\% & 0.95\% & 8.68\% & 12.47\% & 13.69\% & 61.36\% \\
    \bottomrule
    \end{tabular}%
      \label{tab:Transitions_COMMON_after}%
    }
\end{table}%

\begin{table}[t!]
     \centering
     \caption{{\it Common} users: 3-state transition matrices $\Pmat^{3}_{\textit{before}}$,  $\Pmat^{3}_{\textit{after}}$.} 
                \label{tab:Transitions_COMMON_2domains_3states}%
     \resizebox{\columnwidth}{!}{
       \begin{tabular}{lrrrm{0.5em}lrrr}
       \multicolumn{4}{c}{$\Pmat^{3}_{\textit{before}}$} & & \multicolumn{4}{c}{$\Pmat^{3}_{\textit{after}}$}\\
       \toprule
       & \multicolumn{1}{c}{\textbf{GC}} &
        \multicolumn{1}{c}{\textbf{BF}} & 
        \multicolumn{1}{c}{\textbf{GR}} & & & \multicolumn{1}{c}{\textbf{GC}} &
                \multicolumn{1}{c}{\textbf{BF}} & 
                \multicolumn{1}{c}{\textbf{GR}}\\
       \midrule
       {\bf GC} & 42.86\% & 22.45\% & 34.69\% & & {\bf GC} & 22.73\% & 33.06\% & 44.21\% \\
       {\bf BF} & 9.12\% & 34.85\% & 56.03\% & & {\bf BF} & 9.15\% & 37.25\% & 53.59\%\\
       {\bf GR} & 2.91\% & 14.54\% & 82.56\% & & {\bf GR} & 4.24\% & 23.60\% & 72.16\% \\
       \bottomrule
       \end{tabular}%
       }
   \end{table}%
   
   \begin{table}[b]
     \centering
     \caption{{\it Common} users: Summary Mobility and Eigenvalue-based Indices for $\Pmat^{3}_{\textit{before}}$ and  $\Pmat^{3}_{\textit{after}}$. The top (bottom) rows correspond to transitions between the 3 high-level (6 expanded) states.}
     \label{tab:commonMobilityIndices}
     \resizebox{\columnwidth}{!}{
     \begin{tabular}{rlllllll}
       \toprule
       & \multicolumn{1}{c}{\bf IR} & \multicolumn{1}{c}{\bf MU} & \multicolumn{1}{c}{\bf MD} & \multicolumn{1}{c}{$\mathbf M_E$} & \multicolumn{1}{c}{$\mathbf M_2$} & \multicolumn{1}{c}{$\mathbf M_D$} & \multicolumn{1}{c}{$\mathbf M_{\textit{SVD}}$} \\ \midrule
   {\bf Before-3} & 0.5342 & 0.3772 & 0.0886 & 0.6987 & 0.5780 & 0.9238 & 0.5333 \\
   {\bf After-3} & 0.4405 & 0.4362 & 0.1233 & 0.8393 & 0.7670 & 0.9794 & 0.6191 \\
       \bottomrule
       {\bf Before-6} & 0.3118 & 0.4988 & 0.1894 & 0.8259 & 0.5051 & $<$1 & 0.7373 \\
       {\bf After-6} & 0.2197 & 0.5713 & 0.2091 & 0.9364 & 0.7042 & 1 & 0.8306 \\
           \bottomrule
     \end{tabular}
     }
   \end{table}
   
   \begin{table}[b!]
     \centering
     {\scriptsize
     \caption{{\it Common} users: Distances of transition matrices from the immobility matrix {$\mathbf I$}. The top (bottom) rows correspond to transitions between the 3 high-level (6 expanded) states.}
     \label{tab:distances}
     \begin{tabular}{rlllll}
       \toprule
       & \multicolumn{1}{c}{$\mathbf L_1$} & \multicolumn{1}{c}{$\mathbf L_2$} & \multicolumn{1}{c}{$\mathbf{SVD}$} & \multicolumn{1}{c}{$\mathbf D_1$} & \multicolumn{1}{c}{$\mathbf D_3$}  \\ \midrule
   {\bf Before-3} & 2.7947 & 1.1386 & 0.4294 & -1.1838 & -0.5739 \\
   {\bf After-3} & 3.3571 & 1.31382 & 0.5174 & -1.3384 & -0.7197 \\
       \bottomrule
       {\bf Before-6} & 8.2586 & 2.0464 & 0.6197 & -5.2088 & -1.6976 \\
       {\bf After-6} & 9.3639 & 2.2484 & 0.6958 & -6.021 & -2.1765 \\
           \bottomrule
     \end{tabular}
     }
   \end{table}

 As we seek to understand the effects of \SH to the opinions people are exposed, as previously, we restrict our analysis to the common users, and create two transition matrices, $\Pmat_{\textit{before}}$ and $\Pmat_{\textit{after}}$ (Tables~\ref{tab:Transitions_COMMON_before} and \ref{tab:Transitions_COMMON_after} resp.). 

We start with the Summary Mobility Indices, as well as the eigenvalue-based indices 
\cite{Prais55,Shorrocks78,SommersC79,JafryS04,KoutrasD13} that quantify the amount of mobility in the system.
 This category includes the eigenvalue $M_E$, second eigenvalue $M_2$, determinant $M_D$, and Singular Value Decomposition $M_{\textit{SVD}}$ indices. A value of $0$ means to total immobility, and a value of $1$ to perfect mobility.
The first observation on Table~\ref{tab:commonMobilityIndices} is that the immobility ratio (IR) decreases after \SH signifying that users transition between different states more often after than before the event. Specifically, the transitions towards extreme gun rights ($MU$) increase more than the transitions towards extreme gun control ($MD$). Thus, overall, the system moves towards extreme stances and mainly exploration of gun rights. The conclusion that the system exhibits more mobility after the event can also be drawn from the eigenvalue-based indices, all of which increase.

The indices described above are used to assess the underlying mobility behaviors in an {\it individual} transition matrix $\Pmat^n$, but {\it not} the similarities between different transition matrices. To compute the latter, we need to introduce the notion of comparison between matrices. The first step towards this goal is to have both matrices at the same base, which is achieved by computing their deviation from a perfectly immobile system described by the identity matrix ${\mathbf \Qmat=\Imat}$.
Among the matrix distances in the literature, we use: 
\begin{itemize*}
\item the L1-norm $||\Pmat-\Qmat||_1 = \sum_i{\sum_j(p_{ij}-q_{ij})}$,  
\item the $L_2$ norm $||\Pmat-\Qmat||_2 = \sqrt{\sum_i{\sum_j(p_{ij}-q_{ij})^2}}$, 
\item the SVD distance $D_{\textit{SVD}} = M_{\textit{SVD}}(\Pmat)- M_{\textit{SVD}}(\Qmat)$, where $M_{\textit{SVD}}$ is the SVD index defined above, and 
\item two ``risk''-adjusted difference indices, $D_1$ and  $D_3$, which have the advantage of detecting the direction of the transition, while weighing proportionally ``close'' and ``far'' transitions by the factor $(i-j)$:
\begin{equation}
D_1(\Pmat, \Qmat) = \sum_i{\sum_j{(i-j)(p_{ij}-q_{ij})}} \nonumber 
\end{equation}
\vspace{-0.4cm}
\begin{equation}
D_3(\Pmat, \Qmat) = \sum_i{\sum_j{(i-j)sign(p_{ij}-q_{ij})(p_{ij}-q_{ij})^2}}. \nonumber
\end{equation}
\end{itemize*}
The distances are given in Table~\ref{tab:distances}. The $L_1$, $L_2$ and SVD distances show that after \SH the users transition between different states more often than before, while the distances of $\Pmat_{\textit{after}}$ from $\mathbf{I}$ are bigger than the distances of $\Pmat_{\textit{before}}$ from $\mathbf{I}$. 
That conclusion is also corroborated by the ``risk''-adjusted difference indices, which also bear the information (negative values) that the majority of transitions are towards gun rights webpages.

\subsection{Are the Balanced Sites Mediators?} 

Intuitively, one would expect the balanced pages to act as mediators among websites of opposing viewpoints. We are interested in determining whether they are structured to encourage this type of consumption. Is the structure of the web graph such that the balanced pages can be used as jumping-off points?
 
To this end, we enumerate the direct transitions among gun control and rights websites. The number of indirect transitions via a balanced webpage is complementary. The percentages of the direct transitions before, and after Sandy Hook 
are given in Table~\ref{tab:bridges}.

 \begin{table}[h!]
   \centering
   {\footnotesize
   \caption{{\it All} users: Percent of direct transitions between gun control and gun rights webpages.}
   \begin{tabular}{rlll}
     \toprule
     & \multicolumn{1}{c}{\bf Before} & \multicolumn{1}{c}{\bf After} & \multicolumn{1}{c}{\bf Overall} \\ \midrule
 {\bf Control $\rightarrow$ Rights} & 91.50\% & 85.40\% & 86.30\% \\
 {\bf Rights $\rightarrow$ Control} & 82.20\% & 86.90\% & 86.00\% \\
     \bottomrule
   \end{tabular}
   \label{tab:bridges}
}
 \end{table}

We observe that the transitions between gun control and rights occur mostly in a direct way, without accessing a balanced page. Moreover, before \SH, the percent of {\it direct} control to rights transitions (91.5\%) is bigger than the percent of rights to control (82.2\%) transitions, while the opposite holds true after \SH. 
In conclusion, we do not see evidence that the balanced web domains serve as bridges between gun control and rights webpages.

\section{Conclusions}
\label{sec:conclusions}

We have examined the browsing behavior of searchers for the controversial and polarizing topic of gun control.  We focused on the influence of a single disruptive and shocking news event about the tragic massive shooting at the \SH Elementary School in December 2012.  By starting from a large corpus of web logs from November and December 2012, we extracted a footprint of user information-seeking behavior on the URLs that are germane to the topic, and followed a multi-step labeling procedure. Our most interesting findings can be summarized as follows:
\begin{itemize*}
\item All in all, people use the web to largely access agreeable information, as signified by the low diversity of labels capturing viewpoints expressed in visited domains.
\item The domains provide a myopic view in the polarizing topic, showing low diversity in the presented stances.
\item When the external event threatens to influence users directly, they explore content outside their filter bubble.
\item Overall, half of the transitions are from gun control to gun rights pages. As for the \SH shootings, they make the system move into extreme stances, and mainly towards content taking an extreme gun rights stance.
\end{itemize*}

   
We believe that the methods and results shared in this paper represent an initial step in the realm of analyzing log data to understand how people navigate webpages on controversial topics. Future directions include devising interventional studies, such as exploring how ranking and presentation procedures might raise the likelihood that users are exposed to a greater diversity of viewpoints, even if the users do not deliberately seek to access other perspectives, and predicting the changes in polarity after major events that shock the informational system.

\hide{
 In this paper we studied the user browsing behavior for the highly controversial topic of gun control and the effect of the news about the shooting event at the \SH Elementary School in December 2012.
  Starting from the whole corpus of web logs from November and December 2012, we first presented the infrastructure we used to extract the URLs that are germane to the topic, as well as the labeling procedure.
   Then, we analyzed the temporal evolution of the browsing patterns, and the classes of the visited websites. We continued by examining the existing theories about the type of information that people access online -- by employing Shannon entropy --, as well as the effects of the shooting to the observed behavior. Finally, we studied the transitions of the users between different websites by using mobility indices, and showed that transitions towards gun rights webpages were more dominant after \SH. 
   
   The work presented in this paper is just a first step towards the analysis of log data to understand how people navigate webpages on specific controversial topics. There are many interesting future directions, such as devising algorithms to ensure that the users get exposed to diverse information especially for controversial topics -- even if they do not deliberately seek it --, and predicting the polarity change after major events that appear as a shock to the informational system. 
}

\bibliographystyle{plain}
\bibliography{BIB/abbrev,BIB/references}

\newpage
\appendix
\section{Categories of Webpages}
\label{app:categories}
{\small
We seek to label every page that is not ``Off-Topic'' or ``Not Accessible''. Thus, we define symmetric and objective categories:
\begin{itemize*}
\item {\bf Purely Factual}:  The page is on-topic, but only presents facts with no obvious interpretation or commentary on politics.  This may include pages that give statistics regarding guns, laws in different locales about guns, or reporting on news events involving guns where no additional commentary is made. 
\item {\bf Discusses Policies and Issues}:  The page is on-topic and discusses gun policies and issues regarding legislation on gun ownership and usage, or ethical and historical justifications for gun control/rights.  This includes pages that discuss how laws have been interpreted for application in court cases, as well as the personal/official pages of politicians, other persons, organizations, and entities whose stance on gun-related policy is well known even if the page does not feature content currently discussing the policy. The pages in this category are further classified into: 

{\bf Extreme Gun Control}: These pages present a view which favors extreme changes to the current gun laws in an area.  This includes viewpoints that support laws banning any private citizen ownership of guns, as well as what would be viewed as major legislation changes relative to a locale that are not as sweeping.  Pages that use derogatory and insulting language toward those supporting gun rights belong to this category.  Discussion forums and blogs where the preponderance of comments support this view, and webpages giving contact information about only anti-gun organizations belong to this category. 

{\bf Moderate Gun Control}:  These pages present views that favor some to moderate changes to the current gun laws in an area.  This includes views that may view private citizen ownership of guns as acceptable with appropriate conditions and limitations, but argue that the current laws are not sufficient in defining these conditions and limitations.  Discussion forums and blogs where the preponderance of comments support this view belong here. 

{\bf Highly Balanced}:  These pages either discuss both sides with no obvious bias, or present a straightforward discussion of how laws and policy have been interpreted in the past.  For example, pages that discuss of court case reasoning involving guns would fall into this category.  Likewise, educational texts that appear to fairly present both sides would also fall in this category. 

{\bf Moderate Gun Rights}: These pages present a view which favors little to no changes to the current gun laws in an area.  This includes viewpoints that generally support private citizen ownership of guns with appropriate conditions and limitations, and argue that the current laws are generally sufficient. 
This includes pages selling guns that likely would be viewed acceptable for private ownership under appropriate limitations by a moderate gun control viewpoint. Discussion forums and blogs where the preponderance of comments support this view belong here.

{\bf Extreme Gun Rights}: These pages present a view which favors no changes to the current gun laws in an area and argue that current laws may be overly restrictive and intrusive.  This includes viewpoints that claim current laws are an intrusion on individual rights and argue for lessening of any current gun control policies.  Pages that use derogatory and insulting language toward those supporting gun control are included in this category.  This includes pages selling guns or providing information on guns limited not only to those guns viewed acceptable for private ownership under appropriate limitations by a moderate gun control viewpoint but also those falling under currently debated or proposed legislative control. Discussion forums and blogs where the preponderance of comments support this view, and webpages giving contact information about only pro-gun organizations  belong to this category.  
\end{itemize*}

}

%
%


%

\end{document}